\documentclass[
reprint,
superscriptaddress,
 amsmath,amssymb,
 aps,
prb,
]{revtex4-2}

\usepackage{graphicx}
\usepackage{dcolumn}% Align table columns on decimal point
\usepackage{bm}% bold math
\usepackage{hyperref}% add hypertext capabilities
\usepackage{soul} %testo barrato, per evidenziare cambiamenti, puo' essere tolto
\usepackage[dvipsnames]{xcolor} %testo colorato, per evidenziare cambiamenti, puo' essere tolto

\begin{document}

\preprint{APS/123-QED}

\title{Ultrafast all-optical manipulation of the charge-density-wave in VTe$_{2}$}

\author{Manuel Tuniz}
%\email{manuel.tuniz@elettra.eu}
\affiliation{Dipartimento di Fisica, Università degli Studi di Trieste, Italy}
 
\author{Davide Soranzio}%
\affiliation{Institute for Quantum Electronics, Eidgen$\ddot o$ssische Technische Hochschule (ETH) Z$\ddot u$rich, 8093 Zurich, Switzerland}
  
\author{Davide Bidoggia}%
\affiliation{Dipartimento di Fisica, Università degli Studi di Trieste, Italy}

\author{Denny Puntel}%
\affiliation{Dipartimento di Fisica, Università degli Studi di Trieste, Italy}
 
\author{Wibke Bronsch}%
\affiliation{Elettra - Sincrotrone Trieste S.C.p.A., Strada Statale 14, km 163.5, Trieste, Italy}
  
\author{Steven L. Johnson}%
\affiliation{Institute for Quantum Electronics, Eidgen$\ddot o$ssische Technische Hochschule (ETH) Z$\ddot u$rich, 8093 Zurich, Switzerland}
  
\author{Maria Peressi}%
\affiliation{Dipartimento di Fisica, Università degli Studi di Trieste, Italy}
  
\author{Fulvio Parmigiani}
\affiliation{Dipartimento di Fisica, Università degli Studi di Trieste, Italy}
\affiliation{International Faculty, University of Cologne, Albertus-Magnus-Platz, 50923 Cologne, Germany}
\affiliation{Elettra - Sincrotrone Trieste S.C.p.A., Italy}

 \author{Federico Cilento}%
 \email{federico.cilento@elettra.eu}
 \affiliation{Elettra - Sincrotrone Trieste S.C.p.A., Strada Statale 14, km 163.5, Trieste, Italy}

\date{\today}

\begin{abstract}

The charge-density wave (CDW) phase in the layered transition-metal dichalcogenide VTe$_{2}$ is strongly coupled to the band inversion involving vanadium and tellurium orbitals. In particular, this coupling leads to a selective disappearance of the Dirac-type states that characterize the normal phase, when the CDW phase sets in. Here, by means of broadband time-resolved optical spectroscopy (TR-OS), we investigate the ultrafast reflectivity changes caused by collective and single particle excitations in the CDW ground state of VTe$_{2}$. Remarkably, our measurements show the presence of two collective (amplitude) modes of the CDW ground state. By applying a double-pulse excitation scheme, we show the possibility to manipulate these modes, demonstrating a more efficient way to control and perturb the CDW phase in VTe$_{2}$.

\end{abstract}

\maketitle

Understanding the interactions among different degrees of freedom turns out to be crucial for the description of many macroscopic quantum phenomena, such as high-temperature superconductivity, colossal magnetoresistance or ferromagnetism \cite{Tokura2017, Basov2017, Eichberger2010}. In charge-density wave (CDW) systems, electrons and phonons cooperatively condensate to form a new symmetry-broken phase below a transition temperature. The low-temperature phase is characterized by the coexistence of a spatial modulation of the conduction electron density and a periodic lattice distortion (PLD). The resulting CDW ground state exhibits new low-energy collective excitations, named amplitude (AM) and phase modes, which correspond to distortions and translations of the modulated charge density \cite{book_gruner}. Due to this strong interplay between electronic and lattice degrees of freedom, CDW systems have been widely studied and lately, the possibility to manipulate their properties, is attracting great interest \cite{Duan2021, Stojchevska_2014, Zong_2018, Maklar2021}.\looseness=-1

\begin{figure*}[t!]
\includegraphics [width=\textwidth ]{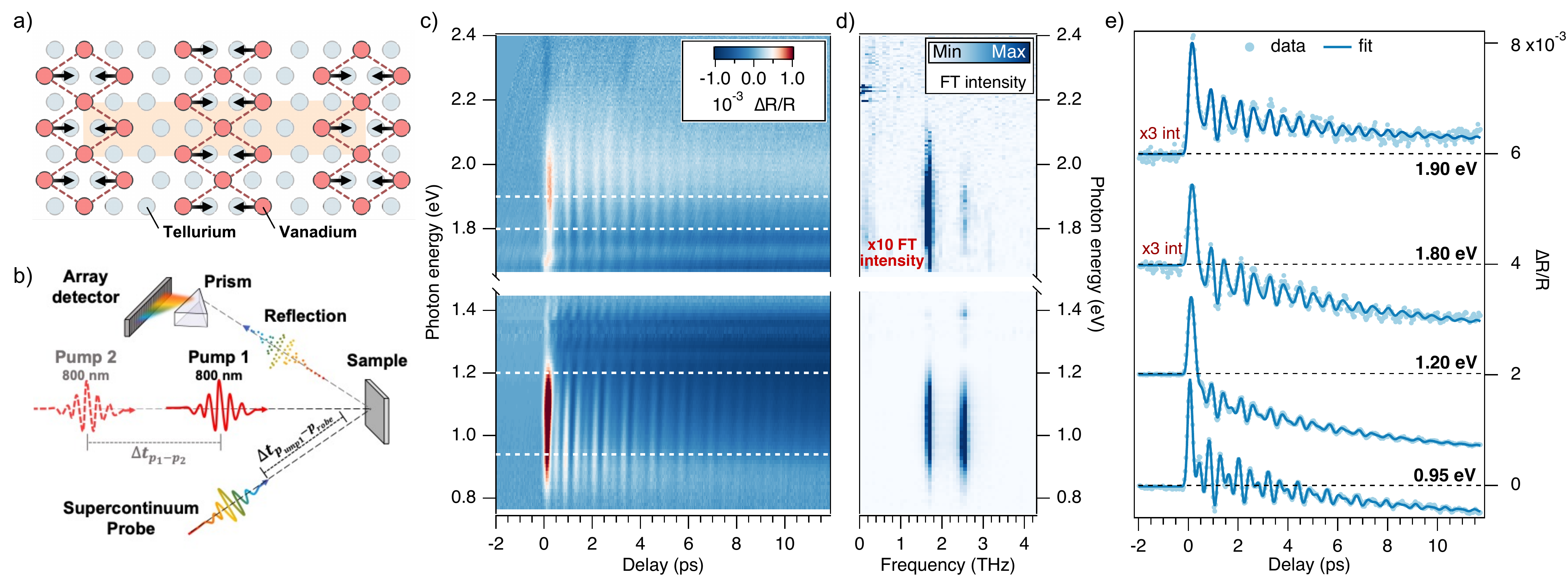}
\caption{\label{fig:1} (a) Top view of a VTe$_{2}$ layer in the low-temperature phase showing the lattice reconstruction that occurs in the CDW phase. The black arrows show the displacement direction of the vanadium atoms while the broken lines highlight the shortest V–V bonds forming the double zigzag structure. The conventional unit cell is indicated by the orange rectangle behind the atoms. (b) Sketch of the time-resolved broadband reflectivity experimental setup. The reflected probe beam is dispersed through a prism and detected by an InGaAs photodiode array (PDA) detector. (c) Two dimensional map showing the evolution of the $\Delta$R/R signal of VTe$_{2}$ as a function of the pump-probe delay and of the probe photon energy. (d) Two dimensional map showing the magnitude of the Fourier Transform (FT) of the coherent part of the $\Delta$R/R signal extracted from (c) as a function of the frequency and of the probe photon energy. For better visibility, in the range between 1.6 and 2.4\,eV the intensity of the FT has been multiplied by a factor 10. (e) Traces extracted from (c) showing the evolution of the $\Delta$R/R signal as a function of the pump-probe delay for four selected probe photon energies.}
\end{figure*}

Transition metal dichalcogenides (TMDCs) are a well known family of layered materials that host a variety of CDWs \cite{Storeck_2020, Stojchevska_2014, Huber_2014, Porer2014}. Among these compounds VTe$_{2}$ is attracting wide interest since recent angle-resolved photoemission experiments have demonstrated the presence of a CDW phase which is strongly coupled to the band inversion involving vanadium 3d and tellurium 5p orbitals. As a consequence, the emergence of the CDW phase leads to a selective disappearance of the topological Dirac-type states that characterize the normal phase\cite{Mitsuishi2020}. The origin of this profound modification of the electronic band structure must be sought in the large lattice reconstruction that breaks the sixfold rotational symmetry of the high-temperature phase. Indeed, in its normal phase VTe$_{2}$ has a CdI$_{2}$ structure, consisting of trigonal layers formed by edge-sharing VTe$_{6}$ octahedra \cite{Bronsema_1984}. Upon cooling, it undergoes a phase transition to the CDW phase at around 475\,K \cite{Ohtani_1981, Bronsema_1984}. The resulting CDW state exhibits a (3$\times$1$\times$1) superstructure characterized by double zig-zag chains of vanadium atoms (shown in Fig.~\ref{fig:1}(a)). Thus, the control of the atomic motion of the atoms involved in the lattice reconstruction is expected to have a profound effect on the electronic modifications induced by the emergence of the CDW phase. In perspective, the manipulation of the CDW in VTe$_{2}$ might result in the possibility to control the topology of the system on ultrafast timescales.

In this work, by means of broadband time-resolved optical spectroscopy (TR-OS), we investigated the ultrafast reflectivity changes caused by collective and single particle excitations in the low-temperature CDW phase of the material. Systematic temperature-dependent measurements show the presence of two phonon modes coupled to the CDW phase. Using a double-pump excitation scheme we show the possibility to manipulate the CDW phase by controlling the displacements of the vanadium atoms involved in the periodic lattice distortion.

Time-resolved reflectivity experiments (Fig.~\ref{fig:1}(b)) were performed using a Ti:sapphire femtosecond laser system, delivering, at a repetition rate of 250\,kHz, $\sim$\,50\,fs light pulses at a wavelength of 800\,nm (1.55\,eV). A broadband (0.75-2.4\,eV) supercontinuum probe beam was generated in a sapphire window. Density functional theory (DFT) simulations were carried out using the \textsc {Quantum Espresso} (QE) \cite{QE-2009, Giannozzi2017,Giannozzi2020} suite of codes. Optimized norm conserving Vanderbilt pseudopotentials \cite{Hamann_2013} with the generalized gradient approximation (GGA) in the Perdew-Burke-Ernzerhof (PBE) parametrization for the exchange-correlation functional \cite{pbe} were employed. A $12\times12\times8$ Monkhorst-Pack k-points mesh was used in sampling the reciprocal space, while cutoff values of 150\,Ry and 450\,Ry were respectively employed for wave-function and charge density representations. Starting from a monoclinic unit cell \cite{Bronsema1984}, our optimized configuration has lattice constants a\,=\,18.984\,\AA, b\,=\,3.595\,\AA, c\,=\,9.069\,\AA{} and $\beta$\,=\,134.62\textdegree{} crystallographic coordinates. To obtain the zone-center phonon eigenvalues and eigenvectors, the dynamical matrix was calculated and diagonalized in a density functional perturbation theory (DFPT) \cite{DFPT2001} approach under the scalar relativistic approximation \cite{Takeda1978} as implemented in QE. \looseness=-1

The two dimensional map reported in Fig.~\ref{fig:1}(c) shows the evolution of the $\Delta$R/R signal as a function of the pump-probe delay in the spectral range accessible through our supercontinuum probe. The measurement was performed at 80\,K, at a fluence of $\sim$490\,$\mu$J/cm$^{2}$ and 1.55\,eV pump photon energy. The spectral region around 1.5\,eV was disturbed by the pump photons scattered from the sample, so it was not considered.

For a wide range of probe photon energies the nonequilibrium reflectivity is dominated by the presence of strong coherent oscillations which add up to the incoherent, exponentially decaying signal. As for other CDW systems, below the critical temperature, the incoherent response of the system shows a first fast decay that vanishes within 1\,ps and a slow recovery process on a time scale of $\sim$10\,ps \cite{Schaefer2013, Schafer_2010, Demsar_1999}. The former, which usually slows down critically upon approaching the critical temperature, is attributed to the recovery of the CDW gap, while the latter is attributed to a second stage of the CDW recovery, as detailed studies of the dynamics as a function of the excitation fluence and applied external electric field have shown in several CDW materials \cite{Eichberger2010,Storeck_2020}. 

\begin{figure}[b!]
\includegraphics [width=\columnwidth ]{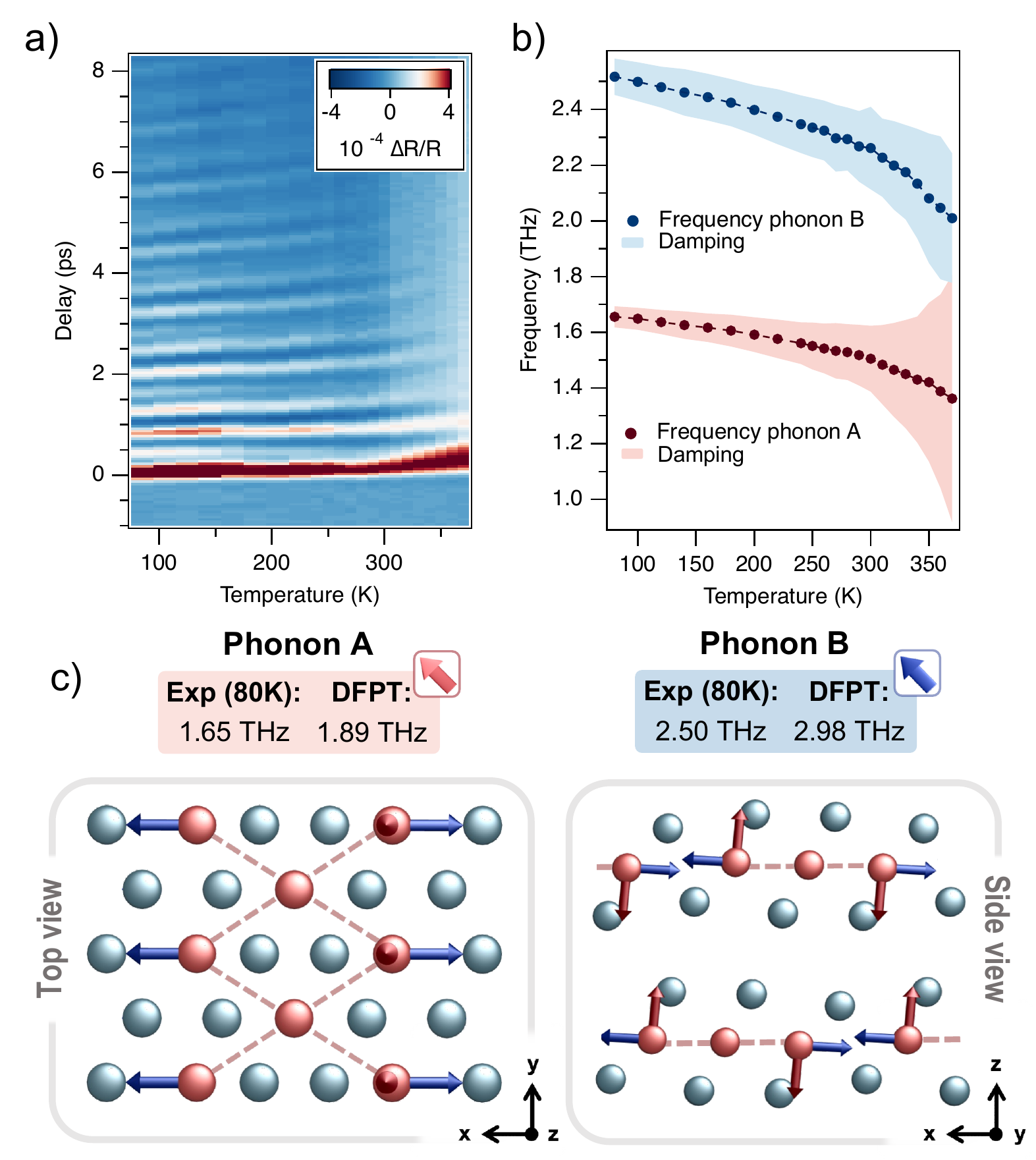}
\caption{\label{fig:2}(a) Two dimensional map showing the evolution of the $\Delta$R/R signal as a function of the temperature and of the pump-probe delay. The measurements have been performed at a probe photon energy of $\sim$\,0.95\,eV (1300\,nm). (b) Evolution of the frequency and the damping (shaded areas whose widths correspond to the damping) of the two phonon modes as a function of the temperature. The renormalization of the frequency follows the behavior expected from a mean field description. (c) Schematic representation of the displacements of the vanadium atoms associated with phonon A (red arrows) and phonon B (blue arrow). Vanadium atoms are in red, tellurium atoms are in teal. The colored boxes show a comparison between the experimental and the calculated frequency of the two phonon modes.}
\end{figure}

Figure \ref{fig:1}(d) shows the square magnitude of the Fourier transform of the coherent part of the signal, isolated by subtracting a double exponential decay fit function to the data shown in Fig.~\ref{fig:1}(c). Two sharp peaks at a frequency of $\sim$1.65\,THz and $\sim$2.50\,THz appear over a wide range of photon energies. As confirmed by our DFPT calculations, these frequencies are linked to A$_{1}$ zone-center optical phonon modes (see the discussion for Fig.~\ref{fig:2}(c)). For notation purposes, since their actual frequency varies with the temperature as we will show in the following, we refer to these modes labelling ``A'' the one with a low temperature frequency of $\sim$1.65\,THz and ``B'' the one with a frequency of $\sim$2.50\,THz.

With the aim of tracking the dependence of both coherent and incoherent parts of the out of equilibrium reflectivity on the probe photon energy, we model the full temporal evolution after the perturbation (time zero, t\,=\,0) as:
\begin{align}
    \frac{\Delta R}{R}(t, h\nu)&= G(t) \otimes \Bigg[ \sum_{i=0}^{2} A_{i}(h \nu) e^{-t/\tau_{i}^{e}(h\nu)} + B(h \nu)  \nonumber \\
    &+ \sum_{j=0}^{2} C_{j}(h \nu) e^{-t/\tau_{j}^{ph}} \cos(\omega_{j}t+ \phi_{j}(h\nu)) \Bigg],
\label{eq:1}
\end{align}
where G(t) represents the cross correlation between the pump and probe pulses. A$_{i}(h\nu)$ denotes the amplitude of the electronic relaxation phenomena with time constant $\tau_{i}^{e}(h\nu)$. B$(h\nu)$ represents the amplitude of much slower process (likely related to the heating of the sample) that in our time window can be approximated by a constant term. Finally, C$_{j}(h\nu)$ denote the amplitude of the oscillation due to the excitation of coherent phonons with angular frequency $\omega_{j}$, phase $\phi_{j}(h\nu)$ and decay time and $\tau_{j}^{ph}$. It is worth noting that the use of a broadband probe offers the unique possibility to study not only the relaxation dynamics of the material but also the effect of the phonon modes on the optical properties of the material over a wide range of energies.

In Fig.~\ref{fig:1}(e) we show four traces extracted at selected photon energies from Fig.~\ref{fig:1}(a) together with the best fits obtained by using Eq.~\ref{eq:1}. For energies below 1.3\,eV a strong beat between the two modes is clearly visible, while for higher energies there is a predominant contribution of phonon A. As shown in the work by Mitsuishi \textit{et al.} \cite{Mitsuishi2020} and confirmed by our density of states calculations, for energies of $\sim$1\,eV the probe beam is resonant with the transition between the hybridized bonding and nonbonding vanadium states above the Fermi level which are strongly affected by the emergence of the CDW phase \cite{supplemental_material}. The fact that for these energies we observe a strong enhancement in the amplitude of the oscillations constitutes the first evidence of the link among these phonon modes and the CDW phase.

\begin{figure*}[t!]
\includegraphics[width=\textwidth ]{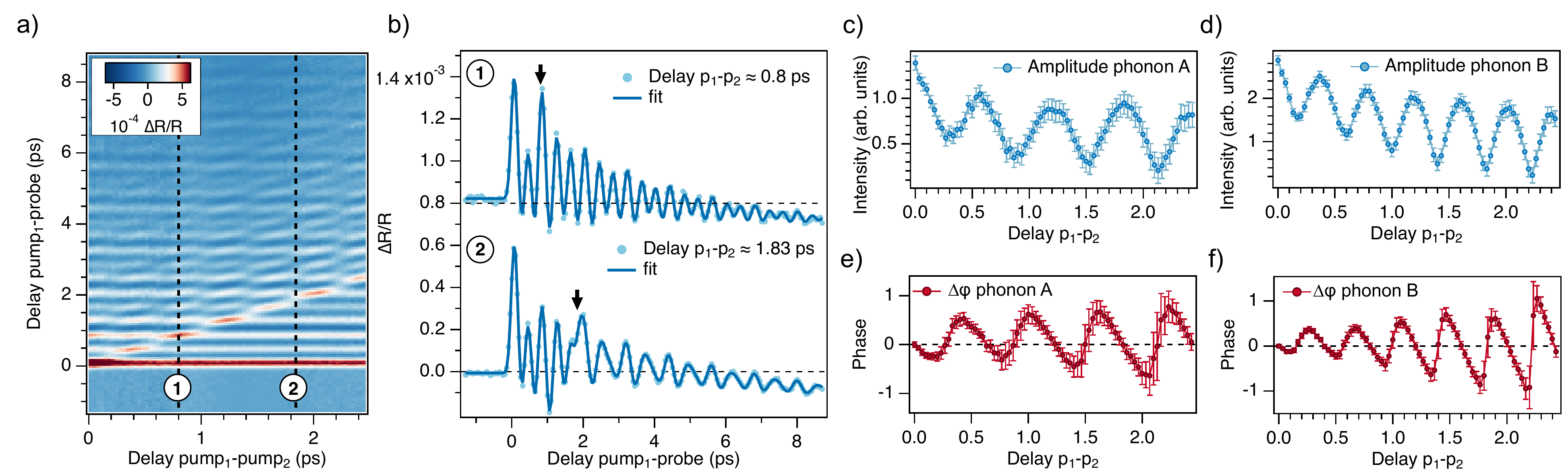}
\caption{\label{fig:3}(a) Two dimensional map showing the evolution of the $\Delta$R/R signal as a function of the delay between the two pump pulses (x-axis) and the delay between the first pump and the probe pulses (y-axis). The measurements have been performed at a probe photon energy of $\sim$\,0.95\,eV (1300\,nm). (b) Traces extracted from (a) together with their best fits showing the evolution of the $\Delta$R/R signal as a function of the delay between the first pump and the probe pulses for two selected delays between the pump pulses \cite{supplemental_material}. The black arrows show the arrival of the second pump pulse. (c-e) Evolution of the amplitude and of the phase shift of the two phonon modes as a function of the delay between the two pump pulses.}
\end{figure*}

Therefore, to study the origin of these phonon modes and, in particular, their interplay with the CDW phase, we performed a systematic study of the evolution of the nonequilibrium reflectivity as a function of the temperature. To this end we focused our attention to the infrared region of the spectrum, where the modulation of the reflectivity due to the two phonon modes is largest. Moreover, to reduce the acquisition time and improve the signal-to-noise ratio we performed single color measurements by filtering the supercontinuum probe beam with a band pass filter centered at $\sim$0.95\,eV (1300\,nm). The two dimensional map in Fig.~\ref{fig:2}(a) shows the evolution of the nonequilibrium reflectivity signal as a function of the temperature while Fig.~\ref{fig:2}(b) shows the renormalization of the frequency and evolution of the damping of the two modes \cite{supplemental_material}. As shown by these two images, a strong renormalization of the frequency of the two phonon modes occurs upon increasing the temperature towards the CDW critical temperature. This effect goes along with an exponential increase of the damping ($\Gamma$\,=\,1/($\pi \tau^{ph})$) of the two modes at high temperatures. This peculiar behavior has been observed in several CDW systems and it is often considered the fingerprint of the amplitude mode of the system \cite{Yoshikawa2021, Chen_2017, Demsar_1999, Yusupov_2008, Kuo_2019, Schafer_2010,Tomeljak_2009}. Remarkably, VTe$_{2}$ shows a rather unusual behavior with two modes strongly intertwined to the CDW phase. This first observation is confirmed by our DFT simulations. Figure \ref{fig:2}(c) shows the eigendisplacements of the two phonon modes for the vanadium atoms involved in the CDW reconstruction. For both phonon modes there is a component of the motion along the direction of the CDW reconstruction, meaning that the two modes can modulate the amplitude of the CDW, as expected from the amplitude mode of the system \cite{Hellmann2012, book_gruner, Sohrt_2014}. Even if both phonons are intertwined with the CDW reconstruction, as shown in Fig.~\ref{fig:2}(c), for phonon B the movement of the vanadium atoms is more localized along the direction of the PLD, therefore this mode is expected to modulate the CDW reconstruction more effectively. This observation is in agreement with the fact that this mode shows a more pronounced temperature renormalization of the frequency, meaning that it is more affected by the CDW disappearance. Conversely, phonon A involves smaller displacements along the lattice reconstruction and more pronounced out of plane movements of the vanadium atoms. We also note that there is reasonable agreement among the calculated and the experimental values of the frequencies, given the fact that the measurements have been performed at 80\,K while the calculations have been performed in the zero temperature limit. We expect to have a further increase of the frequencies at lower temperatures \cite{book_gruner,Schaefer_2014}.\looseness=-1

Coherent control of phonon modes have been demonstrated in a wide variety of crystalline materials including conventional quasi-1D CDW systems \cite{Mihailovic_2002, Gorobtsov2021, Neugebauer_2019, Michael_2022, Nakamura_2019}. Here we apply this approach to the case of VTe$_{2}$, where two phonon modes are directly coupled to the CDW phase.

In general, after photoexcitation, the motion of the vanadium atoms in real space is defined by the combination of the displacements given by the two phonon modes described above. Using a double-pump excitation scheme and tuning the relative delay between the pump pulses, it is possible to selectively enhance or reduce the amplitude of one of two modes and hence it is possible to partially control the atomic motion of the vanadium atoms in real space. We show the results of this approach in the two dimensional map reported in Fig.~\ref{fig:3}(a), where the evolution of the $\Delta$R/R signal is studied as a function of the delay between the two pump pulses. In order to better visualize the effect of the second pump pulse on the oscillations induced by the phonon modes in Fig.~\ref{fig:3}(b) we show two traces extracted from Fig.~\ref{fig:3}(a) at selected delays between the two pump pulses. For this experiment we set F$_{p_{1}}$\,=\,2F$_{p_{2}}$\,$\approx$\,450\,$\mu$J/cm$^{2}$. Therefore, considering the high critical CDW temperature of this material, we reside in the low perturbation regime. 

Trace 1 in Fig.~\ref{fig:3}(b) shows the effect of a second excitation which is in phase with phonon B but almost completely out of phase with phonon A. The result is a strong enhancement of the amplitude of the former and a suppression of the latter. This means that (as shown in Fig.~\ref{fig:2}(c)) the large out of plane component in the movement of vanadium atoms, given by the excitation of phonon A, is deeply reduced while the movement along the coordinate of the PLD is amplified, resulting in a stronger perturbation of the CDW phase. Trace 2 in Fig.~\ref{fig:3}(b) shows the opposite case. Indeed, the excitation of the second pump pulse here is in phase with phonon A but out of phase with phonon B. This leads to an enhancement of the displacements along the direction perpendicular to the CDW reconstruction. Moreover, as demonstrated by Sasaki \textit{et al.} \cite{Sasaki2018}, the double excitation scheme allows us to directly control the phase of the phonon modes \cite{supplemental_material}. Indeed, the second pump pulse can induce a phase shift in the motion of the atoms. As shown by panels (e) and (f) of Fig.~\ref{fig:3}, the extent of the phase shift depends on the delay between the two pump pulses. This effect originates from the fact that, since the two phonon modes are described by damped cosine oscillations, as time passes their intensity decreases (exponentially) while the intensity of the second pump pulse remains fixed. Hence, the effect of the second pump pulse grows with the delay between the two pump pulses. The same behavior is observed in the modulation of the intensity (panels (c) and (d) of Fig.~\ref{fig:3}).

In conclusion, our TR-OS experiments have revealed the presence of two distinct phonon modes coupled the CDW in VTe$_{2}$. As confirmed by our DFT simulations, both modes modulate the amplitude of the lattice reconstruction, hence affecting the charge order. Using a double-pump excitation scheme we have shown the possibility to manipulate the CDW phase by selectively controlling the displacements of the vanadium atoms involved in the periodic lattice distortion. Our finding of more efficient ways to perturb the CDW in VTe$_{2}$ constitutes a first crucial step towards the possibility to control the topology of the electronic structure on ultrafast time scales.\hfill \break

This work was (partially) supported by the Swiss National Science Foundation under project 200020\_192337. 

\providecommand{\noopsort}[1]{}\providecommand{\singleletter}[1]{#1}%

\end{document}